# On Preserving Observation Properties of the Reduced Supervisor in Discrete-Event Systems

Vahid Saeidi[1], Ali A. Afzalian[2*] and Davood Gharavian[3]
Department of Electrical Eng., Abbaspour School of Engineering,
Shahid Beheshti University, Tehran, Iran
[1] v_saeidi@sbu.ac.ir, [2] Afzalian@sbu.ac.ir, [3] d_gharavian@sbu.ac.ir

**Abstract**

Supervisor reduction procedure can be used to construct the reduced supervisor with a reduced number of states in discrete-event systems. The main concepts which are used in this procedure are control consistency of states, control cover, induced supervisor, and normality of the reduced supervisor w.r.t. the original supervisor. In this paper, it is proved that the reduced supervisor, constructed by the proposed method in [9], preserves the observation properties, i.e. normality and relative observability, by self looping corresponding unobservable events at some states of the reduced supervisor. This property can be applied to find a natural projection, under which the supervisor is relative observable.

## 1. Introduction[1]

The state size of a monolithic supervisory controller increases with state sizes of the plant and the specification, as well as the computational complexity [1], and may lead to state explosion [2]. The application of this theory is restricted, and few works are reported on application of this theory in practice, e.g. [3, 4]. Although, modular [5, 6] and incremental [7, 8] approaches try to overcome the complexity of the supervisor synthesis. Other approaches tend to reduce the supervisor for simple implementation. The supervisor reduction procedure, proposed in [9], is an evolution of the proposed method in [10]. This method reduces the redundant information used in supervisor synthesis without any effect on the controlled behavior. A reduced supervisor has some advantages comparing to the main synthesized supervisor, such as simplicity. Although, this procedure is a heuristic method, it has been extended to other applications, e.g. coordination planning for distributed agents [11], and supervisor localization procedure [12].

---

[1] This paper was submitted to the **systems and control letters** in April 19, 2016, for the first time. The revised version of the paper was submitted in September 03, 2016. Currently, it is under review for possible publication.

In practice, engineers want to employ the reduced supervisor instead of the monolithic supervisor. It was proved in the literature [9], that the reduced supervisor is control equivalent to the monolithic supervisor w.r.t. the plant. Whereas, observation properties of the reduced supervisor have not been investigated so far.

In this paper, we show that the observation properties [13-16] of a supervisor can be investigated by self looped events at some states of the reduced supervisor, proposed in [9].

Normality is the strongest observation property of a supervisor, such that the behavior of the supervisor is not affected by some unobserved events. Namely, the synchronization of the projection of a supervisor is equivalent to the supervisor with full observation. We show that, such events which do not affect the behavior of the supervisor are self looped at all states of the reduced supervisor.

Observability and relative observability are other properties of a supervisor (a language, in general), that imply the supervisor can consistently make decision with observation of look-alike strings through the projection channel.

The relative observability property is stronger than the observability property, i.e. a pair of look-alike strings need not to be in the closure of the supervisor in order to make consistently decision. Whereas, in the observability property, both look-alike strings must be in the closure of the supervisor. In [17], the author proposed a method to construct a feasible supervisor corresponding to a (relative) observable one. Each pair of states in the monolithic supervisor can be considered one state in the feasible supervisor by self looping the unobservable event, which occurs between the pair of states. Having a supervisor, we inspect all look-alike strings in the supervisor, to be (relative) observable, under restriction on projection channel. We prove that, similar to the feasible supervisor, the state changes in the reduced supervisor are caused by observable events, only. To find out whether the original supervisor is observable or relative observable, we can test some strings in the supervisor, using the proposed method in [16].

The supervisor reduction procedure only guarantees the control equivalency between the reduced supervisor and the original supervisor w.r.t. the plant, with full observation. To the best of our knowledge, the control equivalency between the reduced supervisor and the original one, under partial observation, has not been reported, so far.

The main concepts in the supervisor reduction procedure are control consistency of states, control cover, induced supervisor, and normality of the reduced supervisor w.r.t. the original supervisor. We prove that, each pair of states, reachable by look-alike strings in a relative observable supervisor is control consistent. We extend this fact to normal supervisors. It will be proved that, the reduced supervisor preserves the relative observability and normality properties by self looping events, which unobservation of them does not violate the consistency in decision making. This can be employed to find a natural projection, which the supervisor is relatively observable under the corresponding projection channel. In fact, we find the tolerable restrictions in the projection channel of a synthesized supervisor, by supervisor reduction procedure, in order to make consistent decisions. It is a useful property of the supervisor reduction procedure. We can employ this procedure to test the relative observability property of a supervisor. Having a synthesized supervisor, we can reduce it, and find events, which appear only as self loop transitions at some states of the reduced supervisor. This method can be employed to investigate observation properties in local controllers which constructed by supervisor localization procedure.

The rest of the paper is organized as follows: In Section 2, the necessary preliminaries are reviewed. Observation properties of the reduced supervisor are investigated in Section 3. In Section 4, two examples on the supervisory control of transfer line, and a guide way, under partial observation, are given to clarify the proposed concepts and propositions. Finally, concluding remarks and future work are given in Section 5.

## 2. Preliminaries

A discrete-event system (DES) is represented by an automaton $\mathbf{G} = (Q, \Sigma, \delta, q_0, Q_m)$, where $Q$ is a finite set of states, with $q_0 \in Q$ as the initial state and $Q_m \subseteq Q$ being the marker states; $\Sigma$ is a finite set of events ($\sigma$) which is partitioned as a set of controllable events $\Sigma_c$ and a set of uncontrollable events $\Sigma_{uc}$, where $\Sigma = \Sigma_c \dot\cup \Sigma_{uc}$. $\delta$ is a transition mapping $\delta: Q \times \Sigma \to Q$, $\delta(q, \sigma) = q'$ gives the next state $q'$ is reached from $q$ by the occurrence of $\sigma$. $\mathbf{G}$ is discrete-event model of the plant. In this context $\delta(q_0, s)!$ means that $\delta$ is defined for $s$ at $q_0$. $L(\mathbf{G}) := \{s \in \Sigma^* | \delta(q_0, s)!\}$ is the closed behavior of $\mathbf{G}$ and $L_m(\mathbf{G}) := \{s \in L(\mathbf{G}) | \delta(q_0, s) \in Q_m\}$ is the marked behaviour of $\mathbf{G}$ [17, 18].

A set of all control patterns is denoted with $\Gamma = \{\gamma \in Pwr(\Sigma) | \gamma \supseteq \Sigma_{uc}\}$. A supervisory control for $\mathbf{G}$ is any map $V: L(\mathbf{G}) \to \Gamma$, where $V(s)$ represents the set of enabled events after the occurrence of the string $s \in L(\mathbf{G})$. The pair $(\mathbf{G}, V)$ is written $V/\mathbf{G}$, to suggest "$\mathbf{G}$ under the supervision of $V$". A behavioral constraint on $\mathbf{G}$ is given by specification language $E \subseteq \Sigma^*$. Let $K \subseteq L_m(\mathbf{G}) \cap E$ be the supremal controllable sublanguage of $E$ w.r.t. $L(\mathbf{G})$ and $\Sigma_{uc}$, i.e. $K = supC(L_m(\mathbf{G}) \cap E)$ [17]. If $K \neq \emptyset$, it can be shown as a

DES, $\mathbf{SUP} = (X, \Sigma, \xi, x_0, X_m)$, which is the recognizer for $K$. If $\mathbf{G}$ and $\mathbf{E}$ are finite-state DES, then $K$ is regular language. Write $|.|$ for the state size of DES. Then $|\mathbf{SUP}| \leq |\mathbf{G}||\mathbf{E}|$. In applications, engineers want to employ $\mathbf{RSUP}$, which has a fewer number of states (i.e. $|\mathbf{RSUP}| \ll |\mathbf{SUP}|$) and is control equivalent to $\mathbf{SUP}$ w.r.t. $\mathbf{G}$ [9], i.e.

$$L_m(\mathbf{G}) \cap L_m(\mathbf{RSUP}) = L_m(\mathbf{SUP}) \tag{1}$$
$$L(\mathbf{G}) \cap L(\mathbf{RSUP}) = L(\mathbf{SUP}) \tag{2}$$

The natural projection is a mapping $P: \Sigma^* \to \Sigma_0^*$ where (1)$P(\epsilon) := \epsilon$, (2) for $s \in \Sigma^*$, $\sigma \in \Sigma$, $P(s\sigma) := P(s)P(\sigma)$, and (3) $P(\sigma) := \sigma$ if $\sigma \in \Sigma_0$ and $P(\sigma) := \epsilon$ if $\sigma \notin \Sigma_0$. The effect of an arbitrary natural projection $P$ on a string $s \in \Sigma^*$ is to erase the events in $s$ that do not belong to observable events set, $\Sigma_0$. The natural projection $P$ can be extended and denoted with $P: Pwr(\Sigma^*) \to Pwr(\Sigma_0^*)$. For any $L \subseteq \Sigma^*$, $P(L) := \{P(s)|s \in L\}$. The inverse image function of $P$ is denoted with $P^{-1}: Pwr(\Sigma_0^*) \to Pwr(\Sigma^*)$ for any $L \subseteq \Sigma_0^*$, $P^{-1}(L) := \{s \in \Sigma^*|P(s) \in L\}$. The synchronous product of languages $L_1 \subseteq \Sigma_1^*$ and $L_2 \subseteq \Sigma_2^*$ is defined by $L_1 \parallel L_2 = P_1^{-1}(L_1) \cap P_2^{-1}(L_2) \subseteq \Sigma^*$, where $P_i: \Sigma^* \to \Sigma_i^*$, $i = 1,2$ for the union $\Sigma = \Sigma_1 \cup \Sigma_2$ [19].

## 3. Observation properties of the reduced supervisor

A procedure was proposed in [9], to reduce the state size of the supremal supervisor. This method constructs a generator which is control equivalent to the monolithic supervisor w.r.t. the plant. Let $\mathbf{SUP} = (X, \Sigma, \xi, x_0, X_m)$ and define $E: X \to Pwr(\Sigma)$ as $E(x) = \{\sigma \in \Sigma | \xi(x, \sigma)!\}$. $E(x)$ denotes the set of events enabled at state $x$. Next, define $D: X \to Pwr(\Sigma)$ as $D(x) = \{\sigma \in \Sigma | \neg\xi(x, \sigma)! \& (\exists s \in \Sigma^*)[\xi(x_0, s) = x \& \delta(q_0, s\sigma)!]\}$. $D(x)$ is the set of events which are disabled at state $x$. Define $M: X \to \{1,0\}$ according to $M(x) = 1$ iff $x \in X_m$, namely the flag of $M$ determines whether a state is marked in $\mathbf{SUP}$. Also, define $T: X \to \{1,0\}$ according to $T(x) = 1$ iff $(\exists s \in \Sigma^*)\xi(x_0, s) = x \& \delta(q_0, s) \in Q_m$, namely the flag of $T$ determines whether some corresponding state is marked in $\mathbf{G}$. Let $\mathcal{R} \subseteq X \times X$ be the binary relation such that for $x, x' \in X$, $(x, x') \in \mathcal{R}$. $x$ and $x'$ are called control consistent, if

$$E(x) \cap D(x') = E(x') \cap D(x) = \emptyset \tag{3}$$
$$T(x) = T(x') \Rightarrow M(x) = M(x') \tag{4}$$

Informally, a pair of $(x, x')$ is in $\mathcal{R}$ if, by (3), there is no event enabled at $x$ but disabled at $x'$, and by (4), $(x, x')$ are both marked (unmarked) in $\mathbf{SUP}$, provided that they are both marked (unmarked) in $\mathbf{G}$. While $\mathcal{R}$ is reflexive and symmetric, it need not be transitive, consequently it is not an equivalence relation. This fact underlies the next definition. A cover $\mathcal{C} = \{X_i \subseteq X | i \in I\}$ of $X$ is called a control cover on $\mathbf{SUP}$ if [9],

$$(\forall i \in I)X_i \neq \emptyset \land (\forall x, x' \in X_i)(x, x') \in \mathcal{R} \tag{5}$$

$$(\forall i \in I)(\forall \sigma \in \Sigma)(\exists j \in I)[(\forall x \in X_i)\xi(x,\sigma)! \Rightarrow \xi(x,\sigma) \in X_j], \qquad (6)$$

Where, $I$ is an index set.

A control cover $\mathcal{C}$ lumps states of **SUP** into cells $X_i$ ($i \in I$) if they are control consistent. A control cover $\mathcal{C}$ is control congruence if $X_i$ are pairwise disjoint. According to (5), each cell of $\mathcal{C}$ is nonempty and each pair of states in one cell should be consistent. According to (6), all states that can be reached from any states in $X_i$ by one transition $\sigma$ is covered by some $X_j$.

Given control cover $\mathcal{C} = \{X_i \subseteq X | i \in I\}$ on **SUP**, an induced supervisor is constructed as $\mathbf{J} = (I, \Sigma, \kappa, i_0, I_m)$ where $i_0 = $ some $i \in I$ with $x_0 \in X_i$, $I_m = \{i \in I | X_i \cap X_m \neq \emptyset\}$ and $\kappa: I \times \Sigma \to I$ with $\kappa(i, \sigma) = j$ provided, for such choice of $j \in I$,

$$(\exists x \in X_i)\xi(x,\sigma) \in X_j \ \& \ (\forall x' \in X_i)[\xi(x',\sigma)! \Rightarrow \xi(x',\sigma) \in X_j] \qquad (7)$$

Overlapping of some states results that $i_0$ and $\kappa$ may not be uniquely determined, thus **J** may not be unique. If $\mathcal{C}$ is control congruence, then **J** is uniquely determined by $\mathcal{C}$. Generally, **J** is control equivalent to **SUP** w.r.t **G**.

A DES **RSUP** $= (Z, \Sigma, \zeta, z_0, Z_m)$ is normal w.r.t **SUP** if,

$$\begin{aligned}&(i)(\forall z \in Z)(\exists s \in L(\mathbf{SUP}))\zeta(z_0, s) = z \\ &(ii)(\forall z \in Z)(\forall \sigma \in \Sigma)[\zeta(z,\sigma)! \Rightarrow (\exists s \in L(\mathbf{SUP}))[s\sigma \in L(\mathbf{SUP}) \ \& \ \zeta(z_0,s) = z]] \qquad (8)\\ &(iii)(\forall z \in Z_m)(\exists s \in L_m(\mathbf{SUP}))\zeta(z_0, s) = z\end{aligned}$$

Given two generators **RSUP** $= (Z, \Sigma, \zeta, z_0, Z_m)$ and $\mathbf{J} = (I, \Sigma, \kappa, i_0, I_m)$ are DES-isomorphic with isomorphism $\theta$ if there exists a map $\theta: Z \to I$ such that

$$\begin{aligned}&(i) \ \theta: Z \to I \text{ is a bijection}\\ &(ii) \ \theta(z_0) = i_0 \text{ and } \theta(Z_m) = I_m \\ &(iii) \ (\forall z \in Z)(\forall \sigma \in \Sigma)\zeta(z,\sigma)! \Rightarrow [\kappa(\theta(z),\sigma)! \ \& \ \kappa(\theta(z),\sigma) = \theta(\zeta(z,\sigma))] \\ &(iv)(\exists i \in I)(\forall \sigma \in \Sigma)\kappa(i,\sigma)! \Rightarrow [(\exists z \in Z)\zeta(z,\sigma)! \ \& \ \theta(z) = i]\end{aligned} \qquad (9)$$

It was proved in [9], if **SUP** is the supremal supervisor for **G** and **RSUP** is any normal supervisor w.r.t **SUP**, such that it is control equivalent to **SUP** w.r.t **G**, then there exist a control cover $\mathcal{C}$ on **SUP** for which some induced supervisor **J** is DES-isomorphic to **RSUP**. In this section, we prove that observation properties (relative observability and normality) are preserved from the monolithic supervisor to the reduced supervisor.

It was defined in [16], that $K$ is relative observable w.r.t. $\bar{C}$, **G** and $P$ (or simply $\bar{C}$-observable) for $K \subseteq C \subseteq L_m(\mathbf{G})$, where $\bar{K}$ and $\bar{C}$ are prefix closed languages, if for every pair of strings $s, s' \in \Sigma^*$ such that $P(s) = P(s')$, the following two conditions hold,

$$(\forall \sigma \in \Sigma) \ s\sigma \in \bar{K}, s' \in \bar{C}, s'\sigma \in L(\mathbf{G}) \Rightarrow s'\sigma \in \bar{K} \qquad (10)$$

$$s \in K, s' \in \bar{C} \cap L_m(\mathbf{G}) \implies s' \in K \tag{11}$$

In the special case, if $C = K$, then the relative observability property is tighten to the observability property. An observation property called normality was defined in [14], that is stronger than the relative observability. $K$ is said to be normal w.r.t. $(L(\mathbf{G}), P)$, if $P^{-1}P(\bar{K}) \cap L(\mathbf{G}) = \bar{K}$, where $L(\mathbf{G})$ is a prefix closed language and $P$ is a natural projection.

Let $\mathbf{SUP} = (X, \Sigma, \xi, x_0, X_m)$ be the recognizer of $K$, $\Sigma_0 \subseteq \Sigma$ and $P: \Sigma^* \to \Sigma_0^*$ be the natural projection. For $s \in \Sigma^*$, observation of $P(s)$ results in uncertainty as to the state of $\mathbf{SUP}$ given by the "uncertainty set" $U(s) \coloneqq \{\delta(q_0, s')|P(s') = P(s), s \in \Sigma^*\} \subseteq Q$. Uncertainty sets can be used to obtain a recognizer for the projected language $P(K)$. By definition of uncertainty set, each pair of states $x, x' \in X$, reachable by $s, s'$, are control consistent, if there exists a nonblocking supervisor $V$ such that $P(s') = P(s) \implies V(s') = V(s)$. $V$ is called a feasible supervisor [17]. Each pair of states $x, x' \in X$ in the monolithic supervisor can be considered one state in the feasible supervisor by self-looping an unobservable event $\sigma$.

In the following proposition, we prove that, each pair of states of a relative observable supervisor, which has an unobservable event in between, are control consistent.

*Proposition 1:* Let $\mathbf{G}$ be a non-blocking plant, described by closed and marked languages $L(\mathbf{G}), L_m(\mathbf{G}) \subseteq \Sigma^*$, and $\mathbf{SUP} = (X, \Sigma, \xi, x_0, X_m)$ be the supervisor of $\mathbf{G}$. Let $\mathcal{R}$ be a set of pairs of control consistent states, defined by (3), (4). Suppose that $K = L_m(\mathbf{SUP})$ and $K \subseteq C \subseteq L_m(\mathbf{G})$ such that $K$ is $(\bar{C}, \mathbf{G}, P)$-observable. If $x, x' \in X$ be a pair of states, $x = \xi(x_0, s)$, $x' = \xi(x_0, s')$ and $P(s) = P(s')$, then $x, x'$ are control consistent.

*Proof:* Assume that $K$ is $(\bar{C}, \mathbf{G}, P)$-observable. Then, $K$ is $(\mathbf{G}, P)$-observable, and we can write,

$$(\forall \sigma \in \Sigma) \; s\sigma \in \bar{K}, s' \in \bar{K}, s'\sigma \in L(\mathbf{G}) \implies s'\sigma \in \bar{K} \tag{12}$$

$$s \in K, s' \in \bar{K} \cap L_m(\mathbf{G}) \implies s' \in K \tag{13}$$

Assume that $\exists \sigma \in E(x) \cap D(x')$. Since $\sigma \in D(x')$, we can write $(\exists s' \in \Sigma^*)[\xi(x_0, s') = x' \& \delta(q_0, s'\sigma)!]$ and $\neg \xi(x', \sigma)!$. From (12), $\sigma$ cannot be disabled at $x'$. Thus, $E(x) \cap D(x') = \emptyset$. Similarly, $E(x') \cap D(x) = \emptyset$.

Now, assume that $T(x) = T(x')$ and $M(x) = 1$. Since $M(x) = 1$, we can say $T(x) = T(x') = 1$. It means that, $s \in K, s' \in \bar{K} \cap L_m(\mathbf{G})$. Thus, from (13), $M(x') = 1$.

□

*Remark 1:* In general, the reverse of Proposition 1 is not true. It means that, each pair of strings $s, s'$, that reach to control consistent states $x, x' \in X$, may not be look-alike. Moreover, the quantifier "all" is used in the relative observability definition, whereas the quantifier "exist" is employed in the definition of control consistency of states. On the other hand, in definition of relative observability, all enabled transitions at one state, should not be disabled at another state. Whereas, in each pair of control consistent states $x, x' \in X$ there should be at least one transition at $x$ ($x'$) that is not disabled at $x'$ ($x$).

We prove that, if a supervisor is relatively observable, then unobservable events- considered in synthesizing the relative observable supervisor- appear just as self loop transitions in the reduced supervisor.

*Theorem 1:* Let **G** be a non-blocking plant, described by closed and marked languages $L(\mathbf{G}), L_m(\mathbf{G}) \subseteq \Sigma^*$, and $\mathbf{SUP} = (X, \Sigma, \xi, x_0, X_m)$ be the supervisor of **G**. Suppose that $K = L_m(\mathbf{SUP})$ and $K \subseteq C \subseteq L_m(\mathbf{G})$ are such that $K$ is $(\bar{C}, \mathbf{G}, P)$-observable, where $P: \Sigma^* \to \Sigma_0^*$. Let $x = \xi(x_0, s), x' = \xi(x_0, s')$ such that $P(s) = P(s')$, $\beta \in \Sigma - \Sigma_0$, and $x = \xi(x', \beta)$. If $\mathbf{RSUP} = (Z, \Sigma, \zeta, z_0, Z_m)$ is the reduced supervisor, then $\beta$ appears just as a self loop transition in **RSUP**.

*Proof:* Assume that $x, x' \in U(s)$, $P(s) = P(s')$, $s = s'\beta$. Moreover, $y = \xi(x_0, t)$ in which $t \in L(\mathbf{SUP})$, and $\alpha, \sigma \in \Sigma$. A typical setup is shown in Fig. 1.

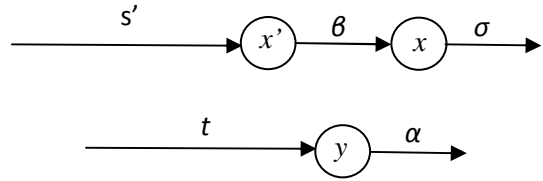

Fig. 1. A set of states and strings in **SUP**

Since $K$ is $(\bar{C}, \mathbf{G}, P)$-observable, $x, x' \in X$ are control consistent, and all enabled events at $x$ are not disabled at $x'$. From the set of states and strings, shown in Fig. 1, we consider two cases to prove that, $\beta$ can just appear as a self loop transition at some states of the reduced supervisor.

<u>Case 1.</u> Assume $(x, y) \in \mathcal{R}$, then $\alpha \notin D(x)$ and $\sigma \notin D(y)$. If $\alpha \in D(x')$, then $(x', y) \notin \mathcal{R}$. Since $\beta \in E(x')$, we can say $\beta \notin D(x)$. It means that $\beta$ may be enabled at $x$ or $s\beta \notin L(\mathbf{G})$. If $\beta \in E(x)$, then $\beta \notin D(y)$. Thus, if $\alpha \notin D(x')$, then we can say $(x', y) \in \mathcal{R}$. In this case, $\beta$ just appears as a self loop transition at lumped states $x, x', y \in X$. The relevant control cover and the corresponding reduced supervisor are shown in Fig. 2.

But, if $s\beta \notin L(\mathbf{G})$, then it is possible that $\beta \in D(y)$, $(x',y) \notin \mathcal{R}$, and the control consistency of $x, y$ is presserved. On the other hand, $s\beta \notin L(\mathbf{G})$ and $\beta \in D(y)$ imply that $\beta$ can not be a transition to reach marked states in the normal reduced supervisor, **RSUP**.

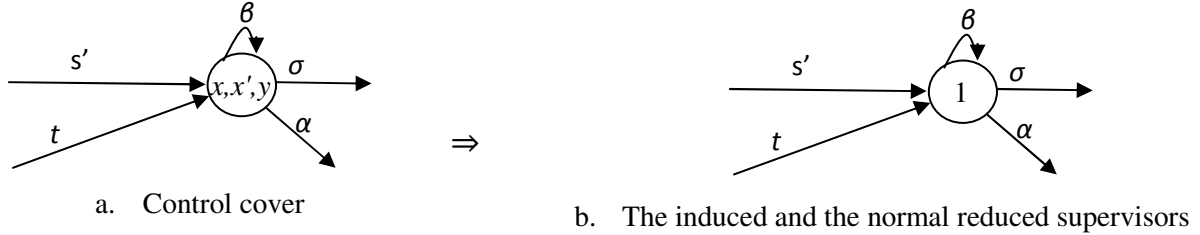

a. Control cover  b. The induced and the normal reduced supervisors

Fig. 2. The control cover and the reduced supervisor in Case 1 ( $x, x', y \in X$ are control consistent)

The relevant control cover and the corresponding normal reduced supervisor are shown in Fig. 3. The lumped states $x, x' \in X$ and $x, y \in X$ are substituted by states 1 and 2, respectively. Since, $\beta$ can not be a transition to reach marked states, thus $\zeta(2, \alpha)$! but $s\alpha \notin L(\mathbf{SUP})$. From $(8 - ii)$, $x'$ must be removed from the state set of the normal reduced supervisor.

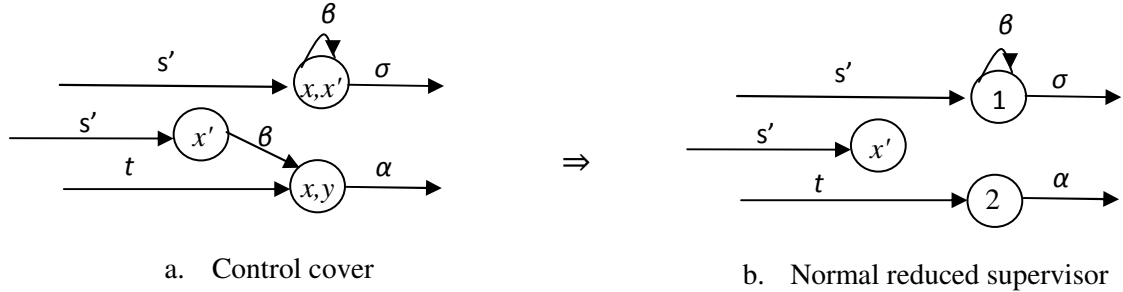

a. Control cover  b. Normal reduced supervisor

Fig. 3. The control cover and the reduced supervisor for **SUP** in Case 1 ( $(x',y) \notin \mathcal{R}$ )

<u>Case 2.</u> Assume $(x', y) \in \mathcal{R}$, then $\alpha \notin D(x')$ and $\beta \notin D(y)$. If $\alpha \notin D(x)$ and $\sigma \notin D(y)$, then $(x, y) \in \mathcal{R}$. In this case, $\beta$ appears just as a self loop transition at the lumped states $x, x', y \in X$ (Fig. 2.a). If $\alpha \in D(x)$ or $\sigma \in D(y)$, then $(x, y) \notin \mathcal{R}$. Hence, the achieved control cover is shown in Fig. 4.a. The lumped states $x', y \in X$ and $x, x' \in X$ are substituted by states 1 and 2, respectively. In order to construct **RSUP** from the control cover, we should consider that each state of **RSUP** must be reachable by a string in **SUP**. Thus, from $(8 - ii)$, we have $[\zeta(2, \beta)!$ and $(s' \in L(\mathbf{SUP}))[s'\beta \in L(\mathbf{SUP}) \& \zeta(z_0, s') = 2]]$. Whereas, reaching $s'\beta$ to state 2, and reaching $s'$ to state 1, do not satisfy $(8 - ii)$. Thus, transition $\beta$ (between states 1 and 2) must be removed, in the construction of the normal reduced supervisor. This is shown in Fig. 4.b.

From aforementioned arguments, $\beta$ can just appear as a self loop transition in some states of the normal reduced supervisor, **RSUP**.

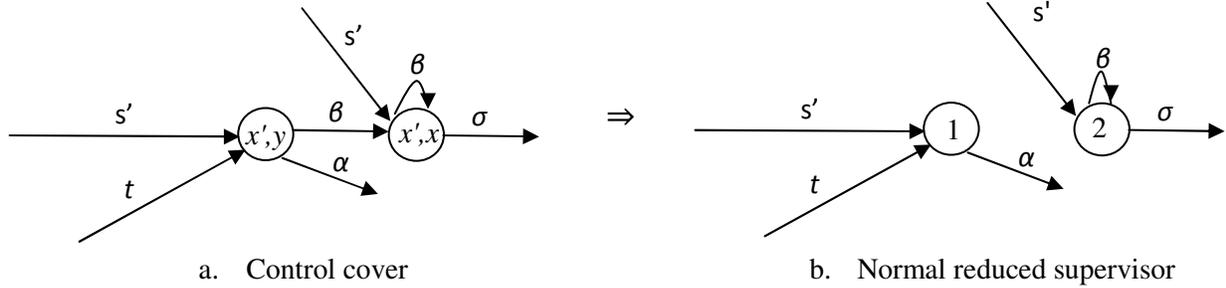

a. Control cover  b. Normal reduced supervisor

Fig.4. The control cover and the reduced supervisor in Case 2 ( $(x, y) \notin \mathcal{R}$ )

□

*Remark 2:* The reverse of Theorem 1 is not true, in general. In other words, there may be some transitions in the reduced supervisor that appear as self loop transitions at some states.

Remark 2 states that, unobservable transitions from one state to other state violates the relative observability of the reduced supervisor. We show this fact in example 4.1.1.

*Corollary 1:* Let **G** be a non-blocking plant, described by closed and marked languages $L(\mathbf{G}), L_m(\mathbf{G}) \subseteq \Sigma^*$, and $\mathbf{SUP} = (X, \Sigma, \xi, x_0, X_m)$ be the supervisor of **G**. Let $\mathbf{RSUP} = (Z, \Sigma, \zeta, z_0, Z_m)$ be a reduced state supervisor. If $K = L_m(\mathbf{SUP})$ is $(\bar{C}, \mathbf{G}, P)$-observable, where $P: \Sigma^* \to \Sigma_0^*$, and $K \subseteq C \subseteq L_m(\mathbf{G})$, then $\mathbf{P(RSUP)}$ is control equivalent to $\mathbf{P(SUP)}$ w.r.t. **G**, i.e.

$$L_m(\mathbf{G}) \cap P^{-1}(L_m(\mathbf{P(RSUP)})) = L_m(\mathbf{G}) \cap P^{-1}(L_m(\mathbf{P(SUP)}))$$
$$L(\mathbf{G}) \cap P^{-1}(L(\mathbf{P(RSUP)})) = L(\mathbf{G}) \cap P^{-1}(L(\mathbf{P(SUP)}))$$

We can achieve another result about finding tolerable restrictions in the projection channel of a supervisor, using self looped events in the reduced supervisor.

*Corollary 2:* If a monolithic supervisor is relatively observable, then the reduced state supervisor presserves the relative observability, by self looping some events at some states of the reduced supervisor. This property can be applied to find a natural projection, such that, the supervisor is relatively observable under the corresponding projection channel. In fact, we can find the tolerable restrictions in the projection channel of a synthesized supervisor, by supervisor reduction procedure.

Corollary 2 declares a useful property of the supervisor reduction procedure. We can employ the supervisor reduction procedure to investigate the relative observability property of a supervisor. Having a synthesized supervisor, we can reduce it, and find events which appear only as self loop transitions at some states of the reduced supervisor. Such events are considered unobservable in the synthesis of the supervisor. Albeit, we may synthesize a relative observable supervisor, not only by employing the procedure proposed in [16], but also by using the supremal controllable sublanguage of a specification. For instance, the synthesized supervisor in example 4.1.1 is relative observable. This property can be employed for investigating observation properties in supervisor localization procedure.

Proposition 1 can be extended to include the normal supervisor.

*Proposition 2:* Let assumptions of Theorem 1 hold true. If $K$ is a normal supervisor w.r.t. $(L(\mathbf{G}), P)$, where $P: \Sigma^* \to \Sigma_0^*$, and $K$ is $L_m(\mathbf{G})$-closed, then

$$(\exists s, s' \in \Sigma^*), x = \xi(x_0, s), x' = \xi(x_0, s'), P(s) = P(s') \Rightarrow (x, x') \in \mathcal{R}$$

*Proof:* Suppose $K$ is a $(L(\mathbf{G}), P)$-normal supervisor. Thus, $P(s\sigma) \in P(\overline{K}), s\sigma \in L(\mathbf{G}) \Rightarrow s\sigma \in \overline{K}$. Assume that $\exists \sigma \in E(x) \cap D(x')$. Since, $\sigma$ is disabled at $x'$, it is observable. Because only observable events can be disabled in a normal supervisor. Hence, $P(s)\sigma \in P(\overline{K})$ and $P(s')\sigma \in P(\overline{K})$. On the other hand, $P(s'\sigma) \in P(\overline{K}), s'\sigma \in L(\mathbf{G}) \Rightarrow s'\sigma \in \overline{K}$. By contradiction, $\sigma$ cannot be disabled at $x'$, and $E(x) \cap D(x') = \emptyset$. Similarly, $E(x') \cap D(x) = \emptyset$.

Suppose $K$ is $L_m(\mathbf{G})$-closed. Namely, $K = \overline{K} \cap L_m(\mathbf{G})$. Assume $T(x) = T(x')$ and $M(x) = 1$. It means that $s \in K$ and $s' \in \overline{K} \cap L_m(\mathbf{G})$. Hence, $s' \in K$, i.e. $M(x') = 1$. Thus, $M(x) = M(x')$. Therefore, $(x, x') \in \mathcal{R}$.

□

We know that a normal supervisor cannot disable an unobservable event $\sigma$. Hence, we conclude, either $\xi(x, \sigma)!$ or $[\neg \xi(x, \sigma)! \Rightarrow (\exists s \in \Sigma^*)(\xi(x_0, s) = x \ \& \ \neg \delta(q_0, s\sigma)!)]$. When $\xi(x, \sigma)!$, from Proposition 2, we have $\exists x' \in X, x' = \xi(x, \sigma)$ such that $x$ and $x'$ are control consistent. In the case of $\neg \xi(x, \sigma)!$, we can make $\sigma$ as a self loop transition at state $x$, because $\neg \delta(q_0, s\sigma)!$. Thus, an unobservable event $\sigma$ is self looped at all states of the reduced supervisor. Since an unobservable event $\sigma$ cannot be disabled at states $x, x' \in X$, $\sigma$ is self looped at $x, x'$, even if they are not control consistent.

## 4. Examples

In this section, we consider examples in order to verify the extended theory in Section 3.

### 4.1. Supervisory control of transfer line with partial observation

Industrial transfer line consisting of two machines $M_1$, $M_2$ and a test unit TU, linked by buffers $B_1$ and $B_2$, is shown in Fig. 5. The capacities of $B_1$ and $B_2$ are assumed to be 3 and 1, respectively. If a work piece is accepted by TU, it is released from the system; if rejected, it is returned to $B_1$ for reprocessing by $M_2$. The specification is based on protecting the $B_1$ and $B_2$ against underflow and overflow [17]. All events involved in the DES model are $\Sigma = \{1,2,3,4,5,6,8\}$, where controllable events are odd-numbered.

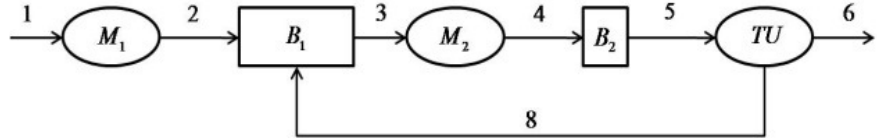

Fig.5. Transfer Line

State transition diagrams of $M_1$, $M_2$, TU and specifications of buffers are displayed in Figs. 6, 7, respectively.

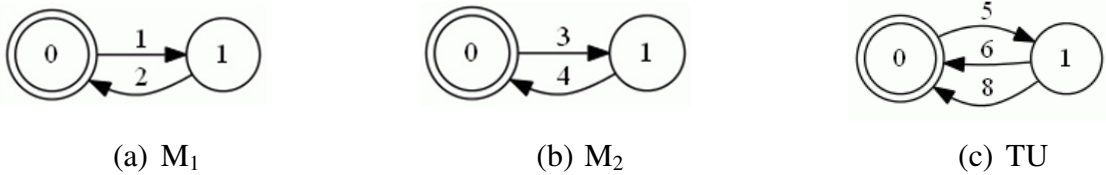

(a) $M_1$          (b) $M_2$          (c) TU

Fig. 6. DES models of M1, M2 and TU

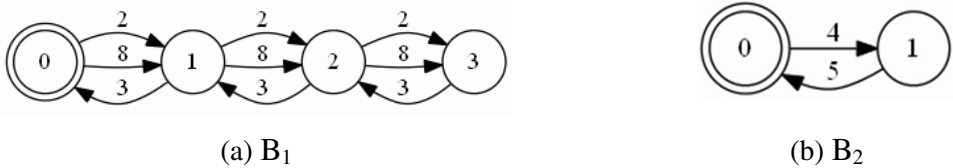

(a) $B_1$          (b) $B_2$

Fig. 7. Specifications of buffers

**4.1.1. The synthesized relative observable supervisor using supcon procedure**

Let the relative observable supervisor is synthesized with $P_0: \Sigma^* \to \Sigma_0^*$, $\Sigma_0 = \Sigma - \{1,3\}$. If we synthesize the supervisor with full observation (i.e. using **supcon** procedure in TCT software [20]), we see that the synthesized supervisor is same as the supremal relative observable supervisor, synthesized by **supconrobs** in TCT software. It means that, a monolithic supervisor can be relatively observable corresponding to the specification and the plant. Using the reduced supervisor (constructed by **supreduce** procedure), we can find events, which can be nulled by the projection channel without violating the consistency in decision making by the supervisor (Fig. 8). Obviously, events 1, 3 appear just as self loop transitions at some states of the reduced supervisor. Although, event 8 is self looped at some states, it also appears as a transition from state 4 to state 1 in **RSUP$_1$**. Thus, the supervisor cannot make consistent decision without observation of event 8.

Suppose $P_8: \Sigma^* \to \Sigma_8^*$, $\Sigma_8 = \Sigma - \{8\}$, and **G** is the DES of the transfer line. Assume an arbitrary ambient language $L(\mathbf{SUP_1}) \subseteq \bar{C}_1 \subseteq L(\mathbf{G})$, such that $s, s' \in \bar{C}_1$, and $P_8(s) = P_8(s')$, in which $s = 1,2,3,4,5,1,8$ and $s' = 1,2,3,4,5,1$. We can write

$s\sigma = 1,2,3,4,5,1,8, \mathbf{3} \in L(\mathbf{SUP_1}), s' = 1,2,3,4,5,1 \in \bar{C}_1, s'\sigma = 1,2,3,4,5,1, \mathbf{3} \in L(\mathbf{G})$

But, we see in Fig. 9, that $s'\sigma \notin L(\mathbf{SUP_1})$. Thus, $L_m(\mathbf{SUP_1})$ is not $(\bar{C}_1, \mathbf{G}, P_8)$-observable.

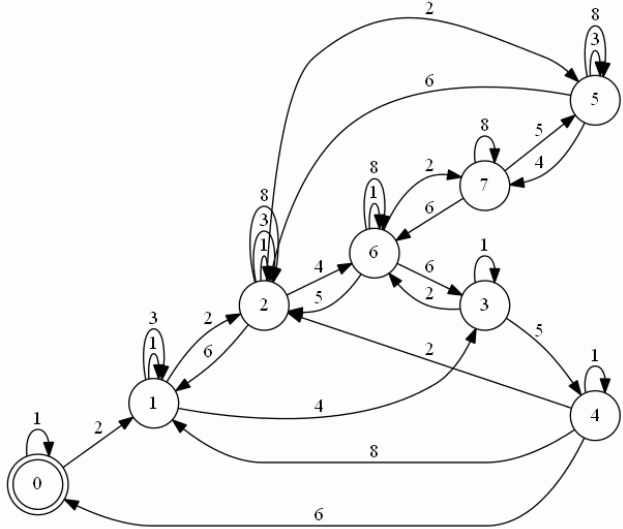

Fig. 8. The reduced state of the relative observable supervisor for transfer line, **RSUP$_1$**

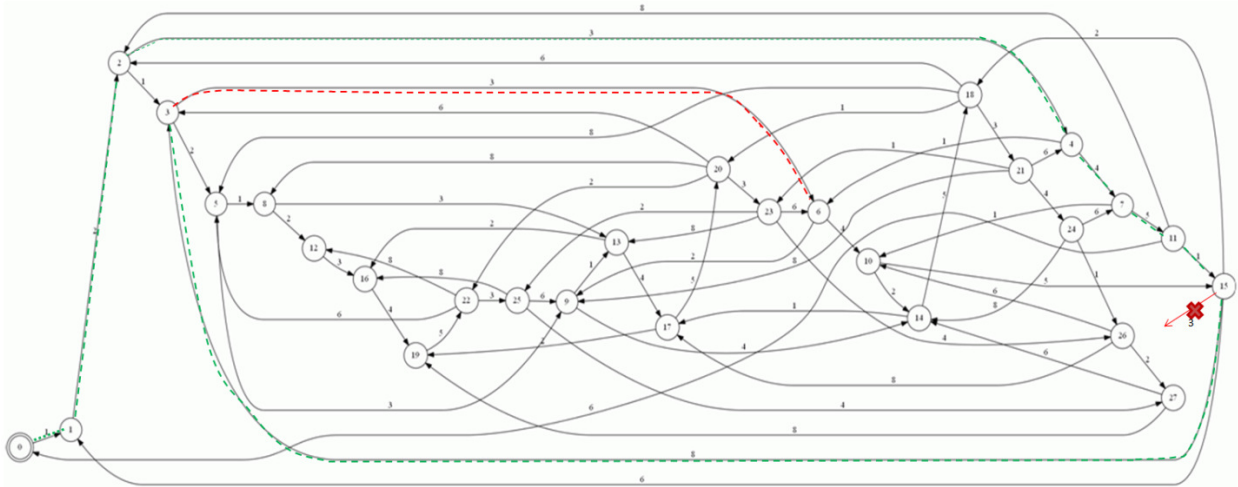

Fig. 9. The relative observable supervisor for transfer line, **SUP$_1$**

### 4.1.2. The synthesized relative observable supervisor using supconrobs procedure

The reduced state supervisor of the synthesized relative observable supervisor, with $P_0: \Sigma^* \to \Sigma_0^*$, $\Sigma_0 = \Sigma - \{1,3,5\}$, is constructed, as shown in Fig. 10. Notice that, the specification is similar to the one given in subsection 4.1.1. We see that, events 1, 3, 5 appear just as self loop transitions, each one at one state of the reduced supervisor (Fig. 10).

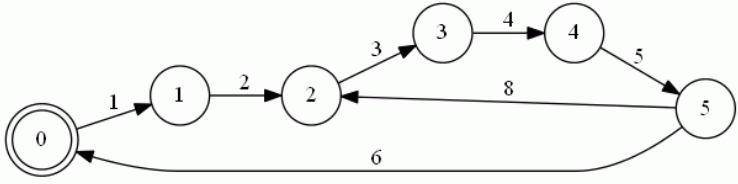
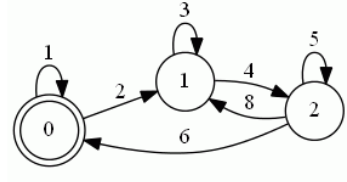

Fig. 10. The supremal relative observable supervisor for transfer line, (**SUP$_2$**)

Fig. 11. The reduced state of the relative observable supervisor for transfer line, (**RSUP$_2$**)

Now, assume that $P_1: \Sigma^* \to \Sigma_1^*$, $\Sigma_1 = \Sigma - \{1\}$. We can find an ambient language $L(\mathbf{SUP_2}) \subseteq \bar{C}_2 \subseteq L(\mathbf{G})$, such that (10), (11) be satisfied, for any pair of look-alike strings (i.e. $P_1(s) = P_1(s')$). For instance, assume that $s\sigma = 1,2,3,4 \in L(\mathbf{SUP_2}), s' = 1,2,3,1 \in \bar{C}_2, s'\sigma = 1,2,3,1,4 \in L(\mathbf{G})$, we see that $s'$ is coming to $L(\mathbf{SUP_2})$ by $\sigma = 4$. Namely, $s'\sigma \in L(\mathbf{SUP_2})$. Also, assume that $s = 1,2,3,4,5,6 \in L_m(\mathbf{SUP_2}), s' = 1,2,3,4,5,1,6 \in \bar{C}_2$. Since $s' = 1,2,3,4,5,1,6 \notin L_m(\mathbf{G})$, (11) is satisfied.

### 4.2. Supervisory control of guide way with partial observation

Consider a guide way with two stations A and B, which are connected by a single one-way track from A to B on a guide way, as shown in Fig. 12. The track consists of 4 sections, with stoplights (*) and detectors (!) installed at various section junctions [17]. Two vehicles $\mathbf{V_1, V_2}$ use the guide way simultaneously. $\mathbf{V}_i, i = 1, 2$ may be in state 0 (at A), state $j$ (while travelling in section $j = 1, \ldots, 4$), or state 5 (at B). The generator of $\mathbf{V}_i, i = 1,2$ are shown in Fig. 13.

The plant to be controlled is $\mathbf{G} = \mathbf{sync}(\mathbf{V_1}, \mathbf{V_2})$. To prevent collision, control of the stoplights must ensure that $\mathbf{V_1}$ and $\mathbf{V_2}$ never travel on the same section of track simultaneously. Namely, $\mathbf{V}_i, i = 1,2$ are mutual exclusion of the state pairs $(i, i)$, $i = 1,\ldots,4$. Controllable events are odd-numbered and the unobservable events 13, 23 are considered to synthesize the supremal relative observable supervisor, i.e. $P_0: \Sigma^* \to \Sigma_0^*$, $\Sigma_0 = \Sigma - \{13, 23\}$. The supremal relative observable supervisor is shown in Fig. 14. The reduced supervisor, in which unobservable events 13, 23 are shown as self loop transitions at state 1, is shown in Fig. 15. Moreover, events 15, 25 are self looped at all states of the reduced supervisor (hence, they are not shown). Thus, the supervisor is normal w.r.t. $(L_m(\mathbf{G}), P_N)$, where $P_N: \Sigma^* \to \Sigma_N^*, \Sigma_N = \Sigma - \{15, 25\}$. Observation of 15, 25 don't affect the control behavior.

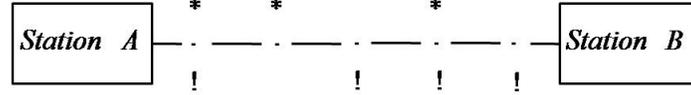
Fig. 12. Schematic of a guide way

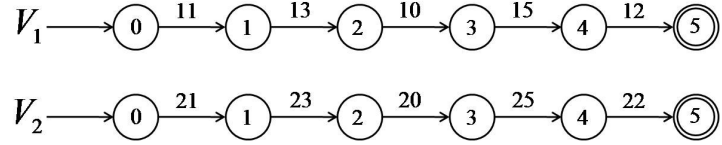
Fig. 13. DES model of each vehicle

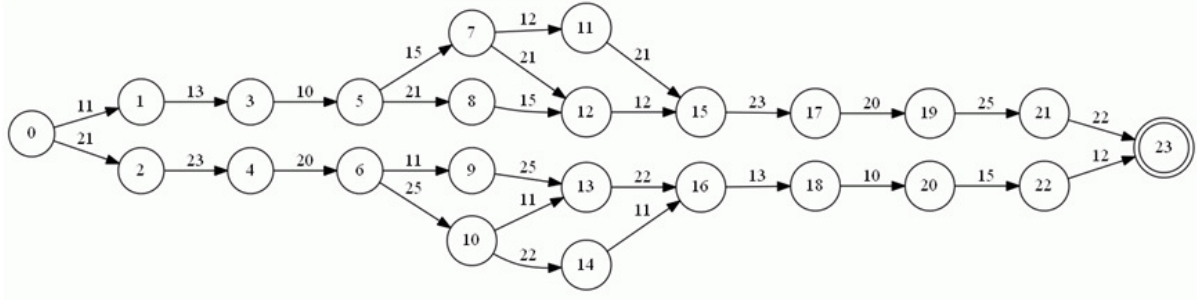
Fig. 14. The relative observable supervisor for the guide way $\Sigma_{uo} = \{13,23\}$

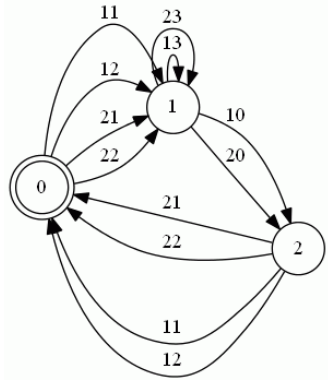
Fig. 15. The reduced state of the relative observable supervisor for the guide way, **RSUP$_3$**

We can find an ambient language $L(\mathbf{SUP_3}) \subseteq \bar{C}_3 \subseteq L(\mathbf{G})$, such that (10), (11) be satisfied, for any pair of look-alike strings, i.e. $P_0(s) = P_0(s')$. For instance,

$$s\sigma = 11,13,10,15,21,12,23,\mathbf{20} \in L(\mathbf{SUP_3}), s' = 11,13,10,15,21,23,12 \in \bar{C}_3,$$

$$s'\sigma = 11,13,10,15,21,23,12,\mathbf{20} \in L(\mathbf{G}) \Longrightarrow s'\sigma \in L(\mathbf{SUP_3})$$

Also,

$$s = 11,13,10,15,21,12,23,20,25,22 \in L_m(\mathbf{SUP_3}),$$

$$s' = 11,13,10,15,21,23,12,20,23,22 \in \bar{C}_3, \ s^{'} = 11,13,10,15,21,23,12,20,23,22 \in$$
$$L_m(\mathbf{G}) \Longrightarrow s' \in L_m(\mathbf{SUP_3})$$

Thus, $L_m(\mathbf{SUP_3})$ is $(\bar{C}_3, \mathbf{G}, P_0)$-observable. We can test in TCT software, that $\mathbf{P_0}(\mathbf{RSUP_3})$ and $\mathbf{P_0}(\mathbf{SUP_3})$ are isomorph. It shows that, Corollary 1 is true.

## 5. Conclusions

This paper addresses preserving the observation properties of the reduced supervisor. We proved that, if a supervisor is relatively observable, then unobservable events appear just as self loop transitions at some states of the reduced supervisor. We showed that, preserving the relative observability in the reduced supervisor by self looping some events can be employed to find a natural projection, which the supervisor is relatively observable under the corresponding projection channel. In fact, we found the tolerable restrictions in the projection channel of a synthesized supervisor, by supervisor reduction procedure, in order to make consistent decisions. This is a useful property of the supervisor reduction procedure. We can employ the supervisor reduction procedure to investigate the relative observability of a supervisor. Having a synthesized supervisor, we can reduce it and find events, which appear only as self loop transitions at some states of the reduced supervisor. The proposed method allows us to use fewer sensors in some cases which safety is not endangered. Moreover, this can be employed for investigating observation properties in supervisor localization procedure. In future work, we will investigate the observation properties of distributed supervisory control, constructed by supervisor localization procedure.